\documentclass[twocolumn,amsmath,amssymb,prl]{revtex4}
\usepackage{latexsym}
\usepackage{graphicx}
\usepackage{psfig}
\usepackage{color}
\usepackage{verbatim}
\usepackage{multirow}
\usepackage[letterpaper, margin=1in]{geometry}
\usepackage{siunitx}
\usepackage[capitalise]{cleveref}

\begin{document}

\preprint{}
\title{Superconducting resonators with voltage-controlled frequency and nonlinearity}
\author{William M. Strickland$^{1}$}
\thanks{These authors contributed equally.}
\author{Bassel Heiba Elfeky$^{1}$}
\thanks{These authors contributed equally.}
\author{Joseph O'Connell Yuan$^{1}$}
\email[Now at Quantum Circuits Inc., New Haven, CT 06511]{}
\author{William F. Schiela$^{1}$}
\author{Peng Yu$^{1}$}
\author{Dylan Langone$^{1}$}
\author{Maxim G. Vavilov$^{2}$}
\author{Vladimir E. Manucharyan$^{3}$}
\email[Now at Institute of Physics, École Polytechnique Fédérale de Lausanne, 1015 Lausanne, Switzerland]{}
\author{Javad~Shabani$^{1}$} \email[Author to whom correspondence should be addressed: ]{jshabani@nyu.edu}

\affiliation{$^{1}$Center for Quantum Information Physics, Department of Physics, New York University, New York 10003, USA}

\affiliation{$^{2}$Department of Physics and Wisconsin Quantum Institute, University of Wisconsin-Madison, Madison, Wisconsin 53706, USA}
\affiliation{$^{3}$Department of Physics, Joint Quantum Institute, and Quantum Materials Center, University of Maryland, College Park, Maryland 20742, USA}

\date{\today}

\begin{abstract}
Voltage-tunable superconductor-semiconductor devices offer a unique platform to realize dynamic tunability in superconducting quantum circuits. By galvanically connecting a gated InAs-Al Josephson junction to a coplanar waveguide resonator, we demonstrate the use of a superconducting element with wideband gate-tunability. We show that the resonant frequency is controlled via a gate-tunable Josephson inductance and that the non-linearity of the InAs-Al junction is non-dissipative as is the case with conventional AlO$_\text{x}$-Al junctions. As the gate voltage is decreased, the inductive participation of the junction increases up to $44\%$, resulting in the resonant frequency being tuned by over \SI{2}{\giga Hz}. Utilizing the wide tunability of the device, we demonstrate that two resonant modes can be adjusted such that they strongly hybridize, exhibiting an avoided level crossing with a coupling strength of \SI{51}{\mega Hz}. Implementing such voltage-tunable resonators is the first step toward realizing wafer-scale continuous voltage control in superconducting circuits for qubit-qubit coupling, quantum-limited amplifiers, and quantum memory platforms.
\end{abstract}

\maketitle

\section{Introduction}

Advances in materials growth, fabrication, and device design have paved the way for the success of superconducting quantum systems based on solid-state platforms~\cite{kjaergaard_currentstateofplay, blais2021, wallraff_strong_2004, blais2004, Koch2007, gyenis2021}. Recently, fixed-frequency superconducting qubits have shown coherence times greater than 1 ms \cite{place2020, somoroff2021}; however, static $ZZ$ crosstalk and parasitic coupling limits two-qubit gate fidelity \cite{mckay2019}. On the other hand, flux-tunable circuits have garnered attention for their ability to eliminate unwanted interactions \cite{yan2018, mundada2019, sung2021, foxen2020}. In addition, they have allowed for fast ($\sim$ \SI{30}{\nano s}) high-fidelity ($>99\%$) two-qubit gates \cite{chen2014}, as opposed to cross-resonance based gates which typically take 150-200 \SI{}{\nano s} \cite{sheldon2016, kandala2021}. These circuits are almost exclusively realized by flux-sensitive superconducting quantum interference devices (SQUIDs) \cite{orlando1999, chiorescu2003,  palacios-laloy_tunable_2008, naaman_-chip_2016, chen2014}. However, conventional flux-tunable circuits have qubit phase coherence limited by low-frequency flux noise \cite{kakuyanagi2007, hutchings_tunable_2017, kumar2016, bialczak2007}. Alternatively, superconductor-semiconductor hybrid structures can be employed to provide voltage tunability for fast and low-power control. The fast gate switching times offered by transistor-like device architectures could be utilized for high-speed coupling and two-qubit gates. In addition, the local control imposed by voltage-tunable devices can offer a flexible design feature for large-scale device footprints, inevitably adding new functionality to the superconducting qubit toolbox.

In a Josephson junction (JJ) with a semiconductor weak link, Cooper pair transport is facilitated by Andreev reflections at the superconductor-semiconductor interface \cite{beenaker1992}. Multiple, coherent reflections form current-carrying Andreev bound states. As the Fermi energy tunes the occupation of each state, such as by an applied gate voltage $V_{G}$, the total critical current $I_{C}$ can be controlled. Such tunability has been routinely demonstrated in current biased devices \cite{mayer2019, Fabrizio17, Henri17, MortenPRA2017, elfeky_local_2021}, and utilized in qubit manipulation, specifically with gatemon qubits \cite{deLange2015, Larsen_PRL, Luthi2018, kringhoj2018, Casparis2018, CasparisBenchmarking, yuan2021, kringhoj_parity2020, danilenko2022, hertel2022}. Furthermore, semiconductor weak-links can directly find applications in the form of couplers \cite{casparis2019, Maxim17, sardashti2020}, amplifiers \cite{phan2022}, superconductor-coupled quantum dot devices \cite{Casparis2018, Scarlino2019, Borjans2020, Burkard2020}, Andreev qubits \cite{Hays2020, hays2021} and nonreciprocal devices \cite{leroux2022}.

In this work, we present a wideband, wafer-scale implementation of a voltage-tunable resonator. The device is based on an InAs two-dimensional electron gas (2DEG) and contacted with epitaxial Al fabricated into a Josephson junction embedded in a coplanar waveguide (CPW) resonator. We show that the Josephson inductance $L_J$ is modified electrostatically by an order of magnitude, allowing for the resonant frequency to be tuned within a \SI{2}{\giga Hz} band. In addition, by studying the high power response of the device we find that the junction exhibits non-dissipative noninearity. The tunability of the resonator frequency allows for strong hybridization with another resonator on the chip. With a maximum coupling strength of $g=$ \SI{51}{\mega Hz}, we achieve strong coupling between the two resonators.

\begin{figure*}
    \centering
    \includegraphics[width=\textwidth]{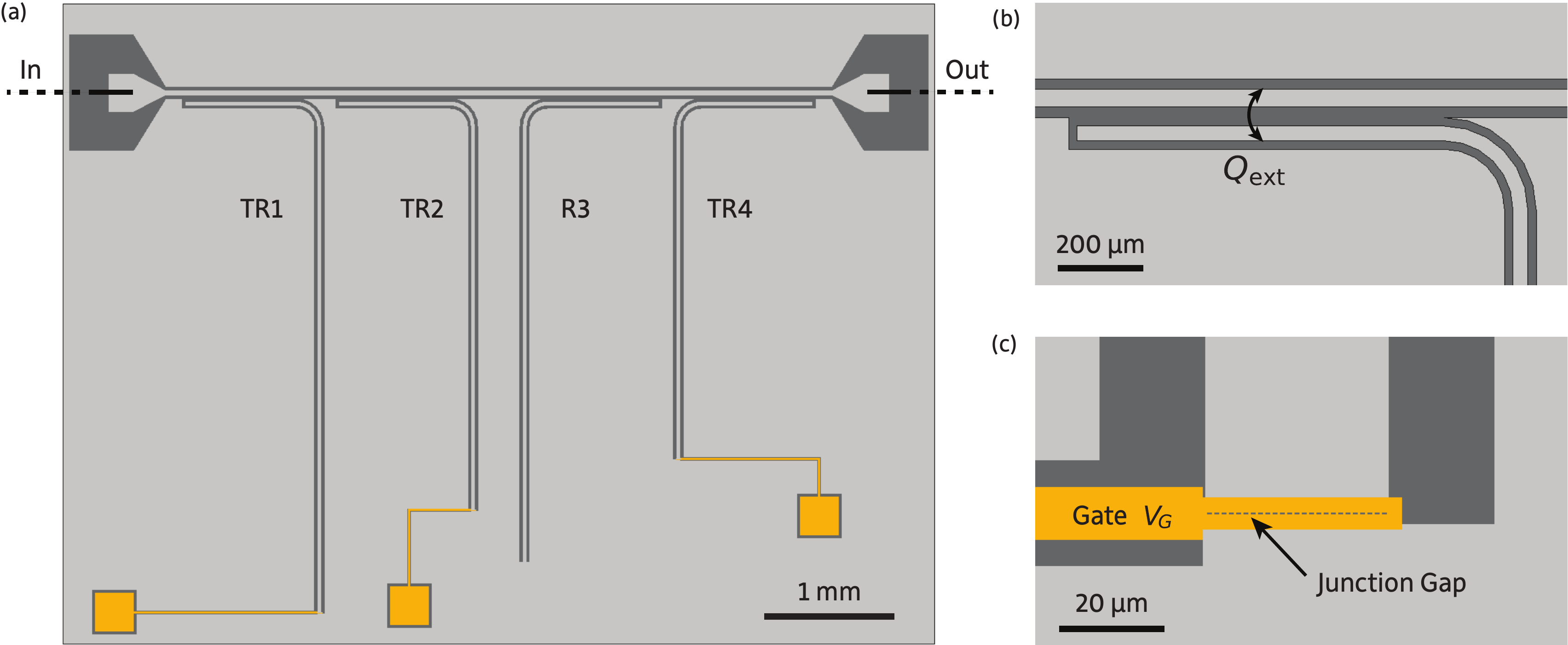}
    \caption{Schematics of the chip design.
    (a) The design consists of four resonators coupled to a common transmission line. Three $\lambda/4$ coplanar waveguides (TR1, TR2 and TR4) are shunted to ground through a Josephson junction and biased by an applied top gate voltage $V_G$. One bare coplanar waveguide (R3) which does not have a junction is used as a reference for kinetic inductance characterization. The inset illustrates the coplanar waveguide geometry layout. (b) The resonators are capacitively coupled to the transmission line with an external quality factor of $Q_\text{ext}$. (c) A closeup illustrating the junction shorting the bottom end of the resonator to ground and the gate. 
    }
    \label{fig:fig1}
\end{figure*}

\section{Device design}

\begin{table}
\centering
\begin{tabular}{|| c | c | c | c | c | c || } 
 \hline
 Resonator & $l$ (mm) & $f_0$ (GHz) & $Q_\text{ext}$ & $C_0$ (pF) & $L_0$ (nH)\\ [0.5ex] 
 \hline\hline
  TR1 & 4.936 & 5.967 & 270 & 0.433 & 1.645 \\ 
  TR2 & 4.136 & 7.111 & 201 & 0.363 & 1.380 \\ 
  R3 & 4.536 & 6.491 & 180 & 0.398 & 1.512\\ 
  TR4 & 3.736 & 7.559 & 152 & 0.341 & 1.298\\ 
  \hline
\end{tabular}
\caption{Design parameters of the coplanar waveguide resonators. The resonant frequency $f_0$ and the external quality factor $Q_\text{ext}$ are calculated by finite element analysis. The capacictance and inductance $C_0$ and $L_0$ are calculated analytically.} 
\label{table:table}
\end{table}

The devices are fabricated on a superconductor-semiconductor heterostructure grown via molecular beam epitaxy \cite{Shabani2016, Kaushini2018, Yuan2020, strickland2022}; details of the growth are discussed in the Supplementary Material. The weak link of the JJ is a high-mobility InAs 2DEG grown near the surface and contacted \textit{in-situ} by a thin aluminum film. The epitaxial heterostructure is grown on a \SI{500}{\micro m} thick InP substrate. We use a III-V wet etch to define the microwave circuit and an Al wet etch to define the JJ. The junction gap is \SI{100}{\nano m} long and \SI{35}{\micro m} wide. We then deposit a \SI{60}{\nano m} AlO$_\text{x}$ gate dielectric, followed by a gate electrode made of layers of Cr and Au which are \SI{5}{} and \SI{50}{\nano m} respectively. A stitched optical image of the wirebonded device is shown in Supplementary Material in Fig. S1.


Measurements are conducted in a dilution refrigerator at a temperature of \SI{30}{\milli K}. A schematic of the measurement setup is shown in the Supplementary Material in Fig. S2, along with further details of the measurement setup. Using a vector network analyzer, we measure the complex transmission coefficient $S_{21}$ as a function of probe frequency $f$. Power is referenced to the output of the vector network analyzer. A method of fitting $S_{21}$ to a circle in the complex plane described in Ref. \citenum{Probst} is used to extract internal and external quality factors, $Q_\text{ext}$ and $Q_\text{int}$, and resonant frequencies $f_{r}$.

A schematic of the chip design is shown in \cref{fig:fig1}(a) with design parameters described in \cref{table:table}. The designs were made using Qiskit Metal~\cite{Qiskit_Metal}. The chip has four CPW resonators with a central conductor width $w = \SI{35}{\micro m}$ and spacing from the ground plane $s = \SI{20}{\micro m}$. This implies a characteristic impedance of $Z_{0} = \SI{48.430}{\ohm}$ calculated using a standard conformal mapping technique \cite{simons_CPW, Pozar2012, goppl2008} assuming a dielectric constant of $\epsilon_{r} = 12.4$ for the InP substrate. The capacitance and inductance of the coplanar waveguides are then given by $C_{0} = \pi / 4 Z_{0} \omega_{0}$ and $L_{0} = 1/\omega_{0}^2 C$ with $\omega_{0 }= 2\pi f_{0}$ where $f_{0}$ is the resonant frequency calculated using finite element analysis simulations \cite{ansys}. These simulations also help us obtain $Q_\text{ext}$, characterizing the coupling to the common feedline.
In three resonators, a Josephson junction is galvanically connected to the end of the CPW, shunting it to ground. We call these devices tunable resonators (TR1, TR2, TR4). The Josephson inductance $L_{J}$ is tunable by an applied gate voltage $V_{G}$ via the top gate. One bare resonator (R3) does not include a shunting Josephson junction and is used as a reference. In this work, we focus on devices TR1, TR2, and R3.

\begin{figure}[ht]
    \centering
    \includegraphics[width=0.45\textwidth]{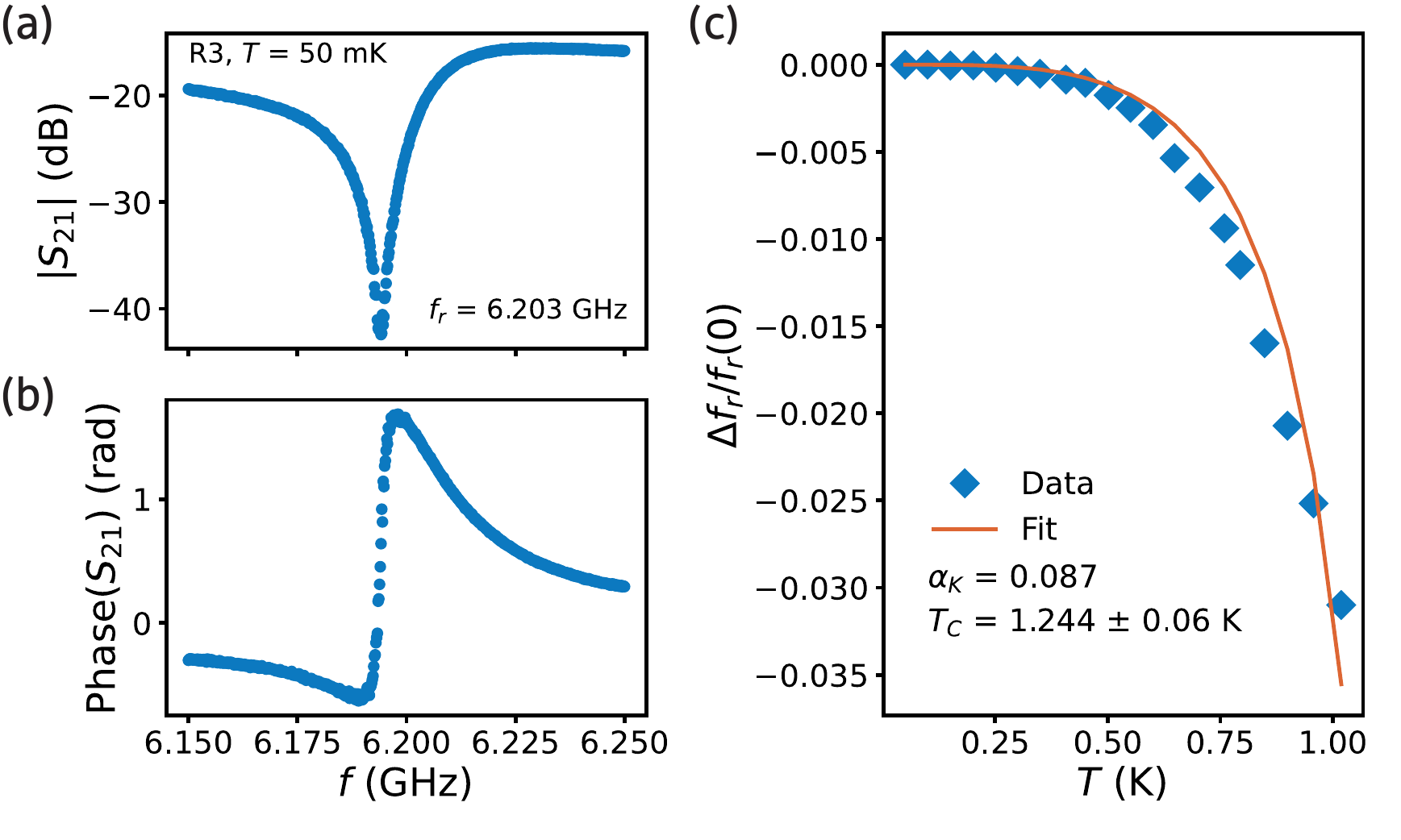}
    \caption{Kinetic inductance extraction from R3. (a) Magnitude and (b) phase of the complex transmission $S_{21}$ shown as a function of frequency. 
    (c) Change in frequency $\delta f_r$ as a function of temperature $T$ as well as a fit to a two-fluids model with $T_C$ as a fitting parameter and $\alpha_K$ fixed. Data was taken at a power corresponding to $\langle n \rangle \sim \SI{3e4}{}$ photons in the cavity.}
    \label{fig:fig2}
\end{figure}

\section{Thin-film kinetic inductance}

In order to accurately determine the inductance contribution of the tunable JJ, we must properly characterize the kinetic inductance of the superconductor thin film. For an Al thickness of $\sim$10 nm, we expect an appreciable kinetic inductance contribution to the total inductance of the resonator. We measure the kinetic inductance fraction $\alpha_K = 1- (f_{r}/f_0)^2$ of a bare CPW where $f_\text{r}$ is the measured frequency of the CPW \cite{gao2006}. \cref{fig:fig2}(a) and (b) shows measurements of the phase and magnitude of complex transmission data $S_{21}$ of the resonant mode of R3. We find the measured resonant frequency to be $f_{r} = \SI{6.204}{\giga Hz}$ while the frequency based on the geometry of the resonator is found to be $f_0 = \SI{6.491}{\giga Hz}$, leading to a kinetic inductance fraction of $\alpha_K = 0.0867$. Considering the geometric inductance of R3, we find a kinetic inductance per square of $L_K^\square = 1.012$ \SI{}{\pico H}. Using a two-fluid model to describe the contribution of kinetic inductance to the total inductance of the CPW \cite{turneaure_surface_1991}, we fit $\Delta f_{r}(T) = f_{r}(T) - f_{r}(0)$ to the equation
\begin{equation}
    \frac{\Delta f_r}{f_r(0)} = - \frac{\alpha_K}{2\left[1-\left(\frac{T}{T_C}\right)^4\right]} + \frac{\alpha_K}{2},
\end{equation}
where $T_C$ is the superconducting critical temperature and $f_{r}(\SI{0}{K}) \approx f_{r}(\SI{50}{\milli K})$. For $\alpha_{K} =  0.0867$, we find that $T_{C}$ = 1.244 $ \pm .060$ \SI{}{\kelvin} corresponding to a superconducting gap of $\Delta_0 = 1.75 k_B T_C =$ 187 $\pm$ 9 \SI{}{\micro eV} consistent with reported values for Al thin films \cite{mayer2019, dartiailh_phase2021}. We note that the kinetic inductance probes the superconducting condensate which also has a contribution from the InAs 2DEG along with the thin film Al \cite{phan2021}.

\section{Gate voltage tunability}

\begin{figure*}[ht!]
    \centering
    \includegraphics[width=0.95\textwidth]{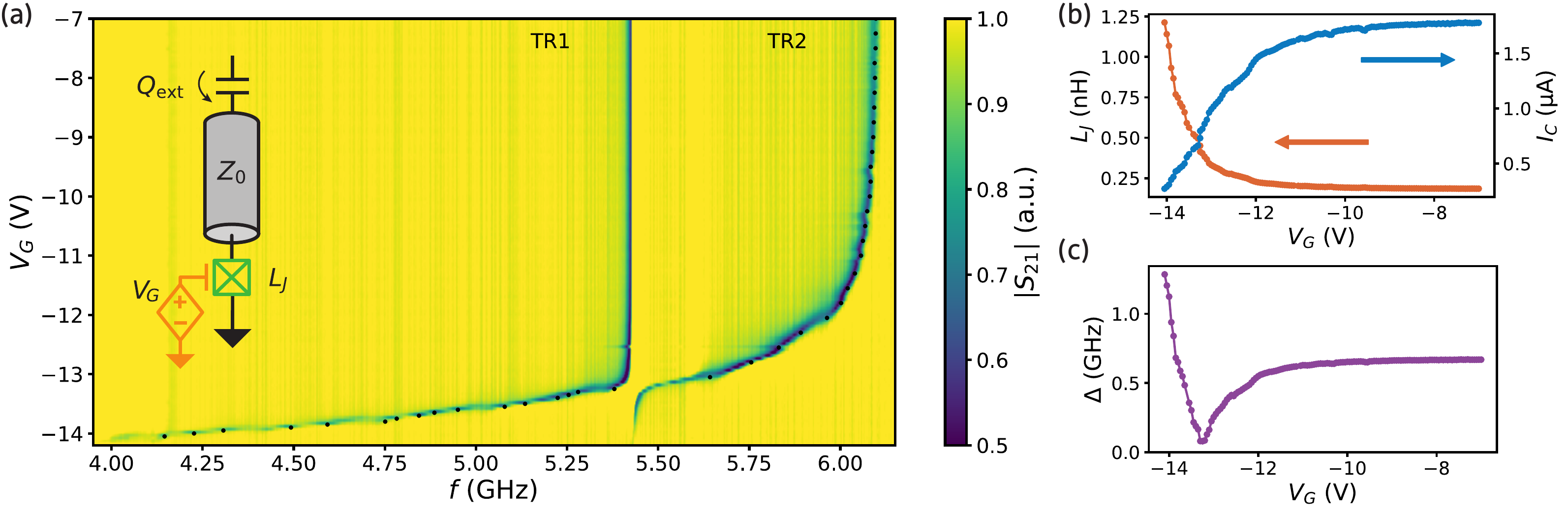}
     \caption{Gate voltage tunability of TR2. (a) The magnitude of the complex transmission coefficient $S_{21}$ as a function of probe frequency $f$ and gate voltage $V_G$, voltage applied to the top gate of TR2. Two resonant modes can be seen, TR1 at 5.4 GHz, and TR2 that starts at 6.1 GHz and is tuned to lower frequencies as $V_G$ decreases. (b) Extracted resonant frequencies of TR2 are mapped to Josephson inductance values $L_J$ of TR2 evaluated from finite-element calculations. Mirrored is the junction critical current $I_C = \Phi_0/2\pi L_J$ where $\Phi_0$ is the magnetic flux quantum. (c) The detuning $\Delta = |f_+-f_-|$ between mode TR1 and TR2 as a function of $V_{G}$.}
    \label{fig:fig3}
\end{figure*}

In order to achieve a wide tunability band, the junction must participate highly in the circuit. By galvanically connecting the junction to the CPW, we create an element that is continuously tuned in a wide dynamic range. In \cref{fig:fig3}(a), we show $|S_{21}|$ versus $f$ while varying the gate voltage $V_{G}$ of TR2. The data taken here is at a power of -96 dBm. Two resonances are observed corresponding to the TR1 and TR2 modes at frequencies $f_1$ and $f_2$, respectively. As $V_{G}$ tunes the occupation of the current carrying conduction channels in the JJ, there is a resulting decrease in the critical current $I_{C}$, which is related to the Josephson inductance through $L_J = \Phi_{0}/2\pi I_{C}$ where $\Phi_{0}$ is the magnetic flux quantum. We find consequently that the TR2 mode is tuned to lower frequencies as the gate voltage decreases, showing a range of \SI{2}{\giga Hz} between V$_{G} = \SI{0}{V}$ and $\SI{-14}{V}$, beyond which the resonant frequency drops outside of the measurement bandwidth of our setup.

Using $L_K^\square$ calculated from R3, we can simulate the effect of the varying Josephson junction $L_{J}$ on the TR2 mode. By accounting for the effect of kinetic inductance in finite element simulations, we calculate the resonant frequency as a function of $L_{J}$, which is represented by a lumped element inductor. Comparing these results to the measured frequencies, we obtain $L_{J}$ and $I_{C}$ as a function of $V_{G}$ shown in \cref{fig:fig3}(b). We find that $L_{J}$ is highly tunable, increasing more than an order of magnitude between the highest and lowest $V_{G}$ points. One can define the Josephson inductive participation ratio in this circuit to be $p_{J} = L_{J}/(L_{J}+ L_{0} + L_{K})$. Using the value of $L_{0}$ for TR2 and $L_{K}$ calculated by the kinetic inductance fraction, we find that at the lowest gate voltage measured $p_{J} = 44.72\%$, implying significant participation of the junction in the circuit. Previous studies based on InAs-Al nanowires have been restricted by either a limited tunability range or discrete switching of the coupler frequency \cite{casparis2019, splitthoff2022}. The wide range and continuous tunability of this 2DEG-based device are advantageous for tunable coupling schemes.

Near $V_G = \SI{-13}{V}$ we find that the two modes undergo an avoided level crossing. We define the difference in frequencies of the two modes as the detuning $\Delta = |f_+-f_-|$ where $f_+$ and $f_-$ are the high and low frequency modes respectively. We show $\Delta$ versus gate voltage in \cref{fig:fig3}(c) and find that at \SI{0}{V} applied, $\Delta = $ \SI{669}{\mega Hz}, while at strongest coupling, $\Delta$ decreases to \SI{79}{\mega Hz}. At large negative gate values, $\Delta$ then increases to \SI{1.285}{\giga Hz} at the lowest frequency of the TR2 mode. The on/off coupling ratio can then be determined as the detuning at the weakest coupling divided by the detuning at the strongest coupling. We find that the on/off coupling ratio at no applied gate voltage with $f_2>f_1$ is 8.47, and at large negative gate voltage with $f_2<f_1$ is 16.27. The latter value is limited by our measurement setup bandwidth and can be expected to increase further. We note that outside the strong coupling regime, the frequency of the TR1 mode remains unchanged due to the local effect of the TR2 gate, resulting in no detectable crosstalk. 

\section{Josephson junction nonlinearity}

\begin{figure*}[ht]
    \centering
    \includegraphics[width=0.9\textwidth]{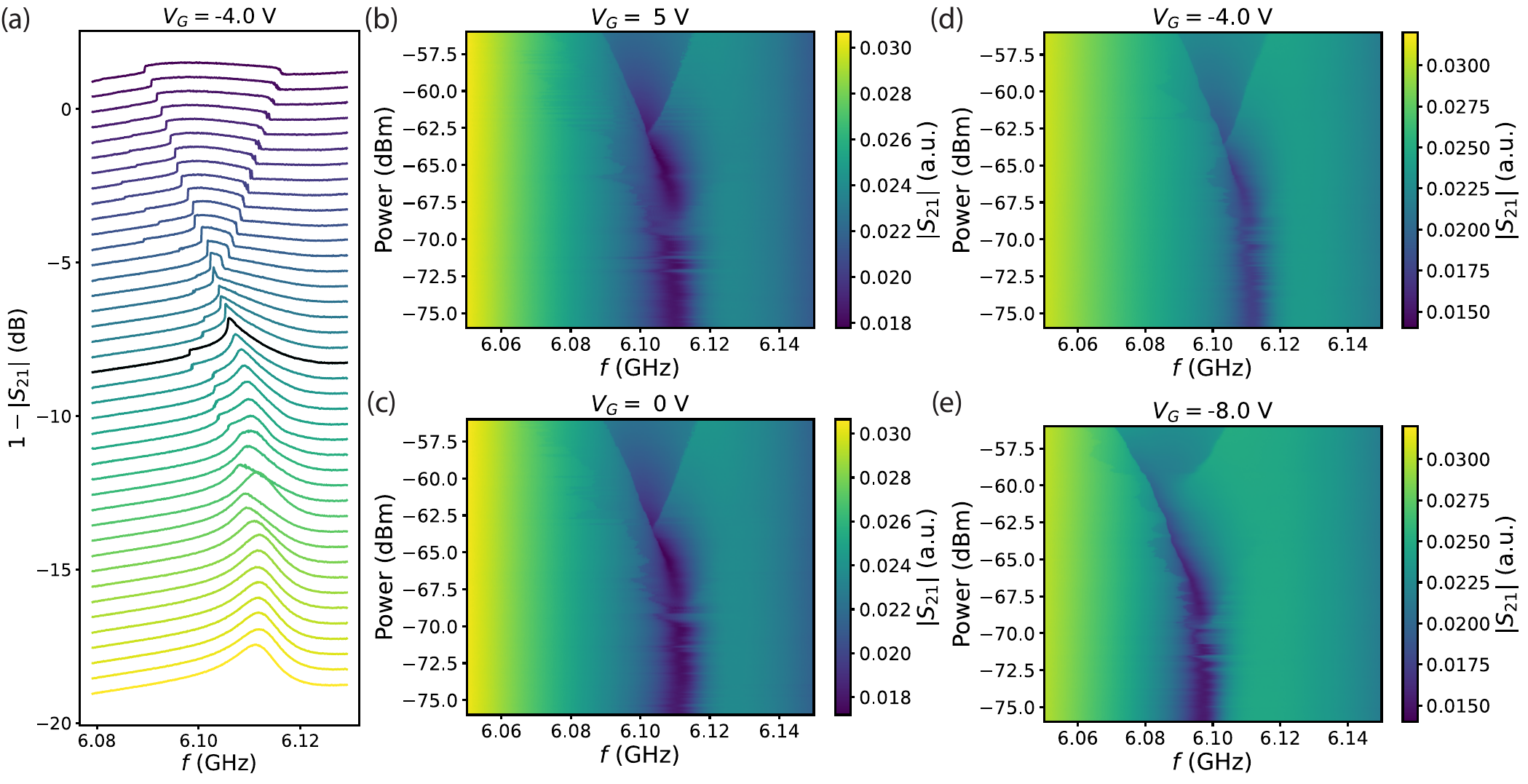}
    \caption{Nonlinear response of TR2. (a) The magnitude of the complex transmission $S_{21}$ plotted versus frequency $f$ at various different input powers, where the top curve is at the highest power of $P=\SI{-56}{dBm}$ while the bottom curve is at $P=\SI{-76}{dBm}$. The curve which corresponds to $S_{21}$ at $P=P_C$ is shown in black.
    (b-e) Power dependence of the Josephson bifurcation at different gate voltages.}
    \label{fig:fig4}
\end{figure*}

\begin{figure}[ht]
    \centering
    \includegraphics[width=0.45\textwidth]{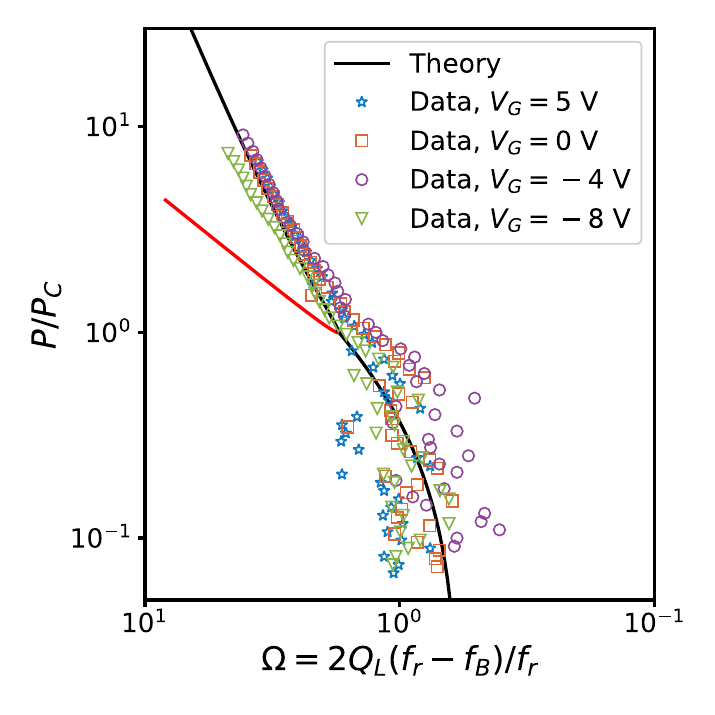}
    \caption{Frequencies at the which the susceptibility $-\partial S_{21}/\partial f$ is maximum, plotted in terms of reduced parameters $\Omega$ and $P/P_C$ for different gate voltages. Solutions to the Duffing equation are shown in black and red, with the two solution branches meeting at $P=P_C$. When plotted in these reduced units, we find there is good agreement between the data and the universal curve.}
    \label{fig:fig6}
\end{figure}

One outstanding question in using voltage tunable junctions is whether the gate voltage control introduces power-dependent dissipation to the system along with the nonlinearity. This has previously been studied by embedding a JJ in a CPW and studying its high power response \cite{boaknin2007}. We employ a similar study as in Josephson tunnel junctions to understand the impact of gate voltage on the junction nonlinearity. 

A combined CPW-JJ system can be described classically by a Duffing oscillator \cite{Landau1976Mechanics, manucharyan2007}, where the nonlinearity gives rise to a power-dependent resonant frequency. At low input powers $P$, the response has a familiar Lorentzian lineshape centered around the resonant frequency $f_{r}$ with a linewidth $Q_{L}$. As $P$ increases, the resonant frequency shifts to lower values, and at a critical power $P_{C}$, the response becomes multivalued with two metastable solutions existing at a single frequency. This phenomenon is known as bifurcation and is the basis for Josephson bifurcation and parametric amplifiers\cite{yurke1989, HoEom2012, siddiqi2004, vijay2009, phan2022}.

In order to keep track of the frequency at which bifurcation develops, we look at the susceptibility $- \partial |S_{21}|/\partial f$. We label the frequency at which the susceptibility is maximum $f_B$. Assuming the low damping limit following Refs. \cite{manucharyan2007} and \cite{boaknin2007}, one can rescale $f_B$ to the reduced frequency $\Omega = 2Q_L(f_r-f_B)/f_r$. In this way, we expect $\Omega$ to be described by the curve
\begin{equation}
P/P_C = \frac{1}{12\sqrt{3}}\Omega^3 \left[1+\frac{9}{\Omega^2} \mp \left(1-\frac{3}{\Omega^2}\right)^{2/3}\right],
\end{equation}
for powers $P>P_C$ and
\begin{equation}
    P/P_C = \Omega\sqrt{3}/2 -1/2,
\end{equation} 
for $P<P_C$.

\cref{fig:fig4}(a) shows the power-dependent response of the TR2 mode with a gate voltage $V_{G} = \SI{-4}{V}$ applied to the gate electrode of TR2. We keep all other gates grounded. As the power increases, the resonant frequency shifts towards lower values, and when $P>P_C$, an apparent discontinuity appears in the data. This is due to the hysteretic behavior of the bifurcation. Since we sweep the frequency in the positive direction, we probe only one solution branch at a time at powers greater than $P_C$, and the apparent discontinuity corresponds to a jump from low to high amplitude solution branches. We find $P_{C}$ by identifying when the susceptibility first diverges. Further details on the procedure used to extract $P_C$ are described in the Supplementary Material and shown in Fig. S5. For $V_{G} = \SI{-4}{V}$, we find that $P_{C} \sim $ \SI{-66}{dBm}; the curve at this power is shown in black in \cref{fig:fig4}(a). We note that the bifurcation of the TR2 mode is absent in the R3 mode. This implies that the nonlinearity in TR2 is mainly caused by the presence of the Josephson junction and not by the kinetic inductance of the thin Al film, as has been reported in other high kinetic inductance superconducting materials \cite{Niepce2019, grunhaupt2018, bretzsullivan2022}.

We analyze the power-dependent response at different gate voltages shown in \cref{fig:fig4}(b-e). We find that as the gate voltage is decreased, $P_{C}$ decreases as expected. Plotted in terms of the reduced frequency, in \cref{fig:fig6} we show $\Omega$ versus $P/P_C$ at four different gate voltage values plotted with the theoretical curves predicted by Eqns. 2 and 3. To rescale $f_B$ to the reduced frequency $\Omega$, $Q_L$ and $f_r$ are extracted from the fit of the resonance at $P=\SI{-76}{dBm}$. A summary of $f_r$, $Q_L$ and $P_C$ used to rescale the data are shown in the Supplementary Material Table S1.

We find that when plotted in terms of the reduced parameters, the data fall on the theoretical curve. Since the solutions to the Duffing model assume the low damping limit, agreement between the data and the theoretical prediction implies that the nonlinearity present in the superconductor-semiconductor junction is not caused by nonlinear dissipative effects. Furthermore, we find that this is true at all four gate voltages, reassuring the fact that applying a gate voltage introduces no additional power-dependent dissipation to the junction. Similar results showing the absence of power-dependent dissipation has been previously reported in AlO$_\text{x}$-Al junctions \cite{boaknin2007}, indicating that InAs-Al junctions have a similar nonlinearity. A discussion about microwave loss mechanisms and mitigation is provided in the Supplementary Material with $Q_\text{int}$ measurements shown in Fig. S3. Deviations from the expected theoretical curve are most likely due to underestimating $f_r$ by taking it at a relatively high power $P=\SI{-76}{dBm}$ compared to the critical powers here.

\begin{figure*}[ht]
    \centering
    \includegraphics[width=0.95\textwidth]{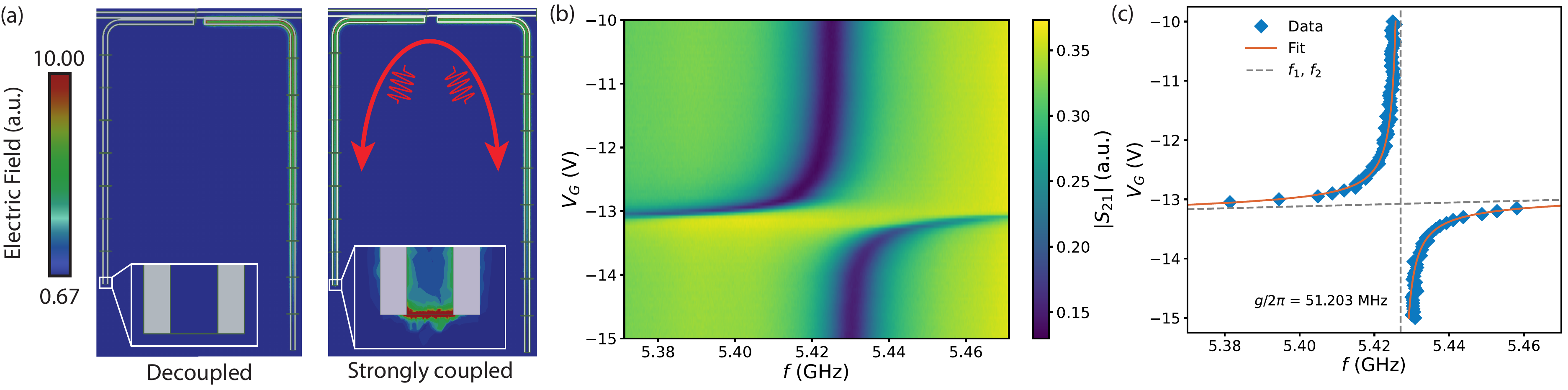}
    \caption{Avoided level crossing of TR1 and TR2 via the gate voltage tuning of TR2. (a) Finite-element calculations of the electric field distribution at a frequency of 5.967 GHz are shown at two values of Josephson inductance. In the ``decoupled'' scenario, the two modes are detuned, where the Josephson inductance is $L_J = 0.00 \SI{}{nH}$, corresponding to \SI{0}{V} applied gate voltage. In the ``strongly coupled'' scenario, the TR2 mode is brought into resonance with TR1 and the two are able to exchange energy mediated by photon swap. This occurs when the Josephson inductance $L_J = \SI{0.38}{nH}$, corresponding to a large negative applied gate voltage. (b) Color plot of the magnitude of $S_{21}$ versus probe frequency $f$ and gate voltage $V_G$. Near $V_G$ = -13 V, strong coupling occurs between the TR2 and TR1 modes and an avoided level crossing is observed. Data is taken at a power corresponding to $\langle n_\text{photon}\rangle \sim \SI{13}{}$ photons. (c) Extracted frequencies as well as a fit to a two oscillators model. The fit yields a coupling strength of $g/2\pi = 51.203 \pm 1.104 \SI{}{\mega Hz}$.}
    \label{fig:fig5}
\end{figure*}

\section{Avoided level crossing}

The wideband voltage tunability allows for coupling different resonators on the same chip. The coupling mechanism can be understood by studying the electric field distribution in the coupled and decoupled regimes, calculated using finite-element analysis methods. We find that in the decoupled scenario, two tunable resonator modes that are detuned by \SI{1}{\giga Hz} have an electric field squared at the frequency of one mode is distributed in the corresponding resonator as expected, shown in \cref{fig:fig5}(a). This occurs when $L_{J}$ of the tunable resonator is set to $L_{J} = \SI{0.00}{\nano H}$. By tuning $L_{J}$ of the tunable resonator towards the frequency of the lower frequency mode, the two modes begin to hybridize, with the electric field energy now occupying both resonators. This corresponds to the strongly coupled regime, occurring when $L_{J} = \SI{0.38}{\nano H}$ with a negative applied gate voltage.

The data presented in \cref{fig:fig5}(b) shows such hybridization between TR2 and TR1 as TR2 is tuned to $V_{G} = \SI{-13}{ V}$ resulting in an avoided level crossing. By extracting the resonant frequencies of the two modes, we fit the data to a simple two-oscillator picture,
\begin{equation}
    f_{\pm} = \frac{1}{2}(f_{1} + f_{2}) \pm \sqrt{\left(\frac{g}{2\pi}\right)^{2} + \frac{1}{2}(f_{1}-f_{2})^{2}} ,
\end{equation}
where $g/2\pi$ is the coupling strength and $f_1$, and $f_2$ are the uncoupled frequencies corresponding to the TR1 and TR2 modes, respectively. In this narrow gate voltage range, we assume the TR2 mode to have a frequency $f_2$, which is approximately linear with the gate voltage changing at a rate of .628 $\pm$ .033 GHz/V. The TR1 mode is fixed at $f_1 = 5.427 \SI{}{\giga Hz}$. Extracting the frequencies $f_{+}$ and $f_{-}$ from the data, we fit the two curves to obtain a coupling strength $g/2\pi = 51.203 \pm 1.104$ MHz as shown in \cref{fig:fig5}(c). This value for $g/2\pi$ taken from the fit to a two oscillators model is consistent with the conventional definition for $g/2\pi$ being equal to half the minimum detuning, $\min(\Delta)/2$, which yields $g/2\pi = $ 51.765 MHz. We note that the coupling strength at this lower power is a bit larger than that which is found at higher power and shown in Fig. \ref{fig:fig3} of $\min(\Delta)/2 = \SI{39.489}{\mega Hz}$.

We hope to show that this device exhibits the basic necessary working principles to implement a tunable coupler using this architecture. The wideband tunablity of the InAs 2DEG device in particular demonstrates the advantage of using this over InAs nanowire based devices. In addition, the large coupling to another device on the same chip demonstrates a proof-of-principle experiment in which this tunable element can be used to couple different superconducting circuit elements on the same chip. We hope to use this device architecture for various applications which include as a tunable coupler between, say, two fixed frequency transmons or two gatemons. It has been shown that for coupler with 10 \SI{}{\micro s} lifetime, a 50 \SI{}{\nano s} CZ gate can be implemented with a gate fidelity of 99.5\% \cite{Maxim17}. An alternative use-case for this circuit is to implement a quantum memory which utilizes dynamically tunable couplers to access ``storage cavities'' which can store information. More details of such a device can be found in Ref. \citenum{sardashti2020}. We hope to have demonstrated in this report the ability to continuously tune the tunable resonator over a wide band, and in future experiments we would like to test dynamic tunability by sending high frequencies control signals through the gate voltage line. 

\section{Microwave loss measurements}

A necessary consideration for the implementation of materials for superconducting qubit circuits is microwave loss. While semiconductor 2DEGs can offer wideband gate-tunable Josephson junctions (JJs), in this section we try to understand mechanisms which limit coherence in our devices.

\begin{table}
\centering
\begin{tabular}{|| c | c | c | c || } 
 \hline
 Device & Superconductor & Substrate & Gate\\ [0.5ex] 
 \hline\hline
  S1 - TR2 & 10 nm \textit{in-situ} Al & 1 \SI{}{\micro m} Buffer & Cr/Au\\ 
  S1 - R3 &  10 nm \textit{in-situ} Al & 1 \SI{}{\micro m} Buffer & - \\ 
  S2  & 100 nm sputtered Al & InP & - \\ 
  S3 & 10 nm \textit{in-situ} Al & 400 \SI{}{\nano m} Buffer & - \\
  S4 &  10 nm \textit{in-situ} Al & 1 \SI{}{\micro m} Buffer & Al \\
  \hline
\end{tabular}
\caption{Summary of samples measured for microwave loss characterization, with TR2 and R3 samples mentioned earlier in the report. Deposition conditions for samples with \textit{in-situ} Al are nominally identical and yield an approximately 10 nm thick film. The substrate refers to both the 500 \SI{}{\micro m} thick InP substrate, as well as the III-V overlayers, referred to here as ``Buffer''. Tunable resonator samples with a Josephson junction and gate are specified with the gate electrode material, being either a Cr and Au combination or Al.} 
\label{table:2}
\end{table}

We fabricate and measure a series of CPW resonators and obtain internal quality factors $Q_\text{int}$ for each sample. These samples aim to study loss mechanisms introduced by specific device conditions. These include loss due to the substrate, the epitaxial III-V layers, the thin film superconductor, and the gate electrode. Details of the devices are discussed in Table~\ref{table:2} and further information can be found in the Supplementary Material.

\begin{figure}
    \centering
    \includegraphics[width=0.5\textwidth]{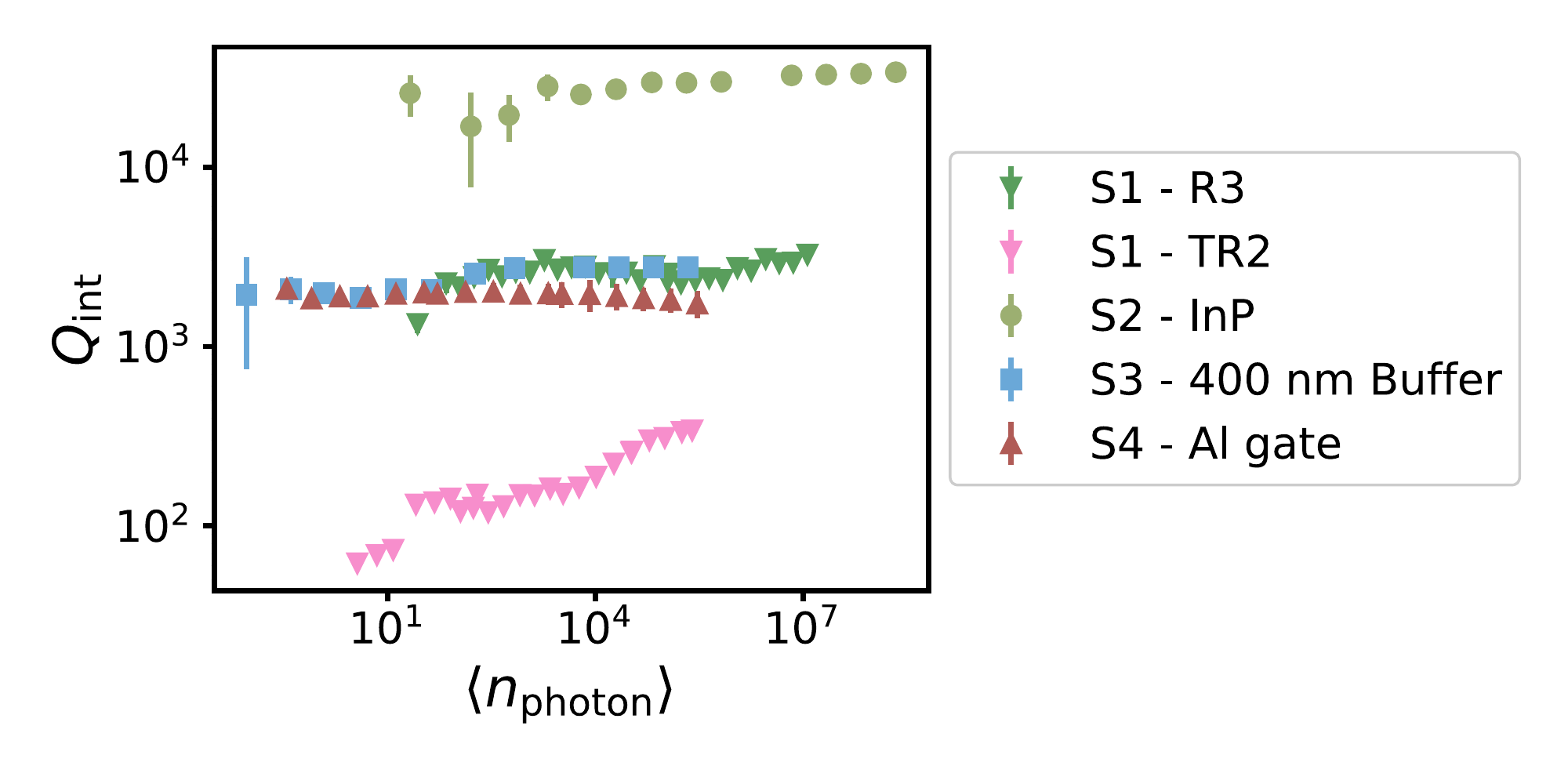}
    \caption{Microwave loss measurements. (a) Internal quality factor $Q_\text{int}$ as a function of average number of photons in the cavity $\langle n_\text{photon} \rangle $ for CPWs on four different samples: a bare CPW, R3, and tunable resonator, TR2, bare CPW on an InP substrate, a bare CPW on a 400 nm Buffer layer, and a tunable resonator with an Al gate electrode.}
    \label{fig:supp1}
\end{figure}

We show a summary of these measurements in \cref{fig:supp1} where we present power dependence of $Q_\text{int}$ for all samples. We find an internal quality factor for the CPW on InP to be $Q_\text{int} = \SI{2.58e4}{} $ at an average photon number of $\langle n_\text{photon} \rangle = 21$. This is consistent with other reports of $Q_\text{int}$ for CPW resonators on InP substrates \cite{Casparis2018} and piezoelectricity has commonly been attributed as the dominant loss mechanism \cite{Scigliuzzo_2020, mcrae2021}. By growing the III-V heterostructure on Si, it should be possible to increase the upper bound on $Q_\text{int}$ for these circuits to more than $\SI{e6}{}$ \cite{mcrae2020}. Alternatively, one can use a flip-chip device design, to concentrate the energy participation in a low loss probe wafer \cite{hazard2022}.

We compare the results for CPWs on the InP substrate to the bare resonator device, R3 on sample S1. We find that at low power, $Q_\text{int} = \SI{2.25e3}{}$. To understand the source of this added loss, we also measure a sample with a 400 nm epitaxial III-V layer, S3. We find that $Q_\text{int}$ is almost identical to that of R3 in sample S1 which has a 1 \SI{}{\micro m} epitaxial III-V layer. By accounting for differences in energy participation of the buffer layers for the \SI{400}{\nano m} and \SI{1}{\micro m} buffers, being 3.5\% and 7.0\% respectively. Details of the participation ratio calculations can be found in the Supplementary Material. Despite the increased epitaxial III-V layer participation, we find that in our devices the epitaxial III-V layers do not significantly affect $Q_\text{in}$. Thus, we believe this decrease in $Q_\text{int}$ is due to the thin film Al. This has previously been observed in other high kinetic inductance materials, such as NbTiN thin films and nanowires \cite{peltonen2016, peltonen_hybrid_2018, bretzsullivan2022}.

We next consider loss due to the gate electrode. While the large participation of the Josephson junction in the circuit provides considerable frequency tunability, the electric field across the junction may couple to the gate electrode. We find that in sample S1, at low power, the TR2 mode has $Q_\text{int} = \SI{1.43e2}{}$ an order of magnitude lower than that of R3 on the same chip. We compare this to sample S2 which has a tunable resonator which replaces the Cr/Au gate with Al. We find that $Q_\text{int}$ of this device is \SI{1.83e3}{} at low power, approaching that of S1 - R3 and S3. This suggests that TR2 has coherence limited by the Cr/Au gate electrode. In addition to the findings raised hee, promising new directions in reducing losses in these materials are using hexagonal boron nitride (h-BN) as a gate dielectric, \cite{Barati2021, Wang2022}. This discussion hopefully provides a path forward for increasing $Q_\text{int}$ of tunable resonators on InAs 2DEGs.

\section{Conclusion}

In conclusion, we have demonstrated the wideband tunability of a superconductor-semiconductor based tunable resonator. We show that the gate voltage tunable junction has a non-dissipative nonlinearity, ideal for implementing in superconducting qubit circuits. After adjusting for the kinetic inductance of the Al thin film, we find that the Josephson inductance is tunable by up to an order of magnitude, achieving a high participation in the circuit of 44$\%$. This high participation enables us to continuously tune the resonant frequency of the tunable resonator mode by more than 2 GHz. We also show that by tuning the tunable resonator mode into resonance with another resonator on the chip, we observe hybridization of the two modes through an avoided level crossing, with coupling strength of \SI{51}{\mega Hz}. The wide tunability range results in large detuning of the two modes resulting in an on/off detuning ratio of $\sim$16 at large negative gate voltage and $\sim$8 at \SI{0}{V} applied. While $Q_\text{int}$ is quite low, we find that the coherence is limited by the normal metal gate line and discuss tangible improvements which can be made to material and device design which can significantly enhance $Q_\text{int}$ of InAs 2DEG tunable resonators. The ability to achieve strong coupling and large detuning between the two modes makes this device design an ideal platform for qubit-qubit coupling schemes and quantum information storage where the TR mode, acting as a coupler, can be brought into resonance with a fixed frequency mode by dynamically pulsing the gate.

\section{Acknowledgements}

We thank Patrick J. Strohbeen, Matthieu Dartiailh, Jaewoo Lee, and Nicholas Materise for fruitful discussions. The authors acknowledge support from the Army Research Office agreement W911NF2110303. The N.Y.U. team acknowledges support from the Army Research Office agreement W911NF2210048 and from the National Science Foundation agreement 2340-206-2014878 A01. W.M.S. acknowledges funding from the ARO/LPS QuaCR Graduate Fellowship. W.F.S. acknowledges funding from the NDSEG Fellowship.  This work was performed in part at the Nanofabrication Facility at the Advanced Science Research Center at The Graduate Center of the City University of New York.

\newpage

\section{Appendix A: Materials growth}
The devices studied are fabricated on a heterostructure grown by molecular beam epitaxy. On an epi-ready, semi-insulating \SI{500}{\micro m} thick InP (100) substrate, a \SI{50}{nm} thick In$_{0.52}$Al$_{0.48}$As/In$_{0.53}$Ga$_{0.47}$As superlattice of ten periods is grown followed by a 50 nm thick In$_{0.52}$Al$_{0.48}$As layer and an \SI{800}{nm} thick In$_{x}$Al$_{1-x}$As graded buffer layer in which the composition is step graded between $x=0.52$ and $0.81$ in steps of $\Delta x = 0.02$. This is followed by a \SI{50}{\nano m} In$_{0.81}$Al$_{0.19}$As virtual substrate. The structure is then modulation doped with Si at a density of $n_D = \SI{1e12}{\centi m^2}$. Following a 6 nm In$_{0.81}$Al$_{0.19}$As spacer, an InAs near-surface quantum well is then grown between two layers of In$_{0.81}$Ga$_{0.19}$As, where the top barrier layer is \SI{10}{\nano m} thick and the bottom barrier layer is \SI{4}{\nano m} thick. The structure is then capped with a 10 nm layer of Al grown \textit{in-situ}. Further details on the materials growth procedure are provided in Refs. \cite{Shabani2016, Kaushini2018, Yuan2020, strickland2022}. Through low temperature magnetotransport measurements, we find the wafer used in this device has a 2D electron density of $n =\SI{9.49e11}{cm^{-2}}$ and an electron mobility of $\mu = \SI{1.45e4}{cm^2/Vs}$ measured along the $[110]$ crystal direction. This corresponds to an electron mean free path of \SI{233}{nm}. With the Josephson junction weak-link being \SI{100}{\nano m} long, the junction is expected to be in the short ballistic regime \cite{mayer2019}. 

\section{Appendix B: Design and fabrication}

\begin{figure}[htpb]
    \centering
    \includegraphics[width=0.45\textwidth]{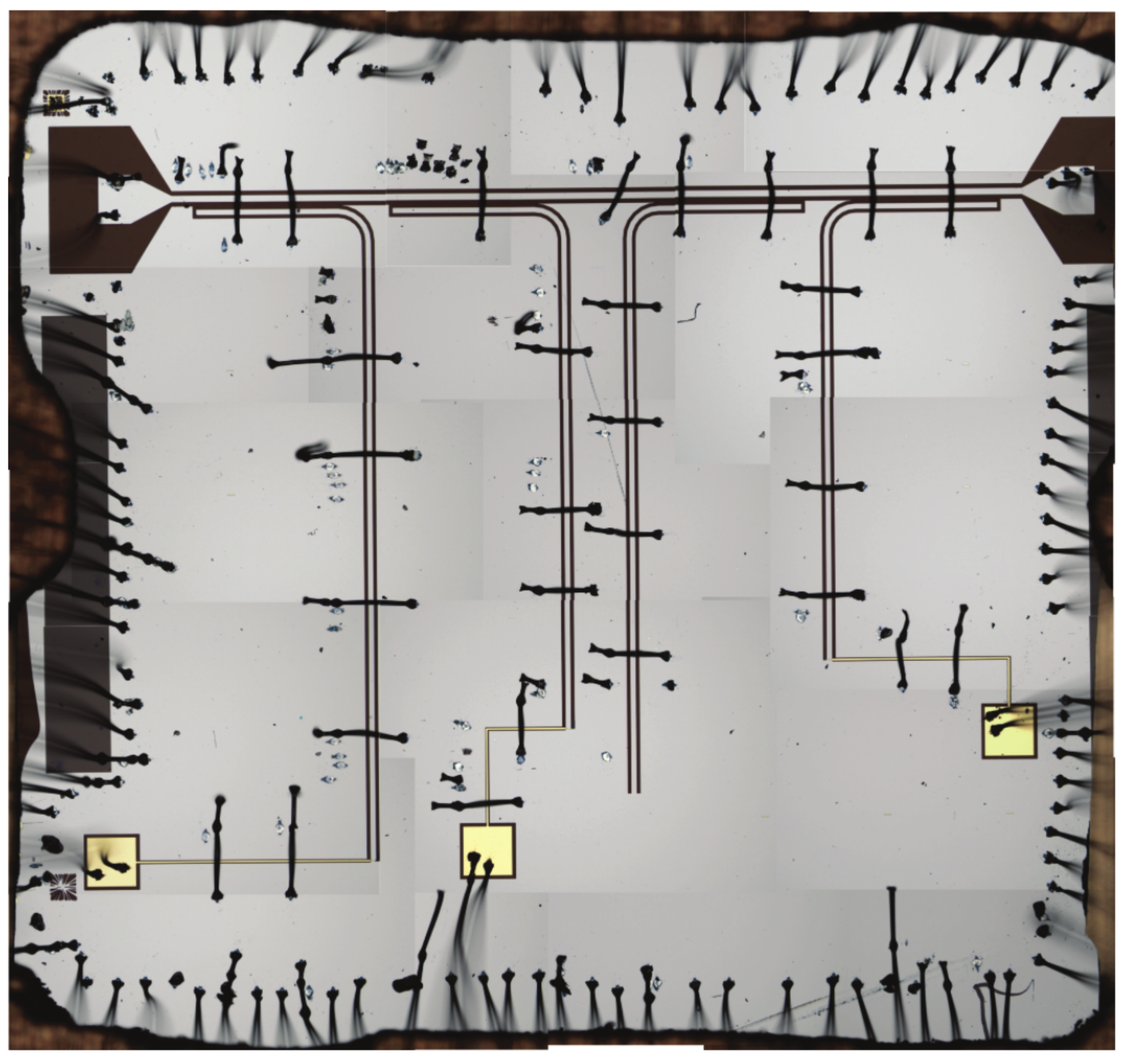}
    \caption{Fabricated chip image. Stitched optical image of the wirebonded device loaded in the microwave sample holder.}
    \label{fig:supp4}
\end{figure}

The design was constructed using Qiskit Metal \cite{Qiskit_Metal} and rendered in Ansys's high frequency simulation software (HFSS) \cite{ansys} to simulate for the expected resonant frequency, external quality factors and electromagnetic field distribution. We use electron beam lithography to define patterns in spin-coated polymethylmethacrylate (PMMA) resist. To define the microwave circuit, Al is removed with Transene Al etchant type-D followed by a wet etch down to the buffer layer using a III-V etchant consisting of phosphoric acid (H$_3$PO$_4$, 85\%), hydrogen peroxide (H$_2$O$_2$, 30\%) and deionized water in a volumetric ratio of 1:1:40. To define the Josephson junctions, we etch away a $ \SI{100}{\nano m}$ long, \SI{35}{\micro m} wide strip of aluminum from the CPW mesa. We then deposit \SI{60}{\nano m} of AlO$_\text{x}$ to serve as a gate dielectric by atomic layer deposition at \SI{120}{\celsius}. This is followed by another electron beam lithography patterning step to define the gate pattern and an electron beam deposition of the gate electrode, consisting of \SI{8}{\nano m} Cr and \SI{80}{\nano m} of Au. An optical image of the fabricated and wirebonded chip is shown in \cref{fig:supp4}. 

A series of samples are used to fabricate CPW devices for $Q_\text{int}$ measurements. The first is a CPW with 100 nm Al deposited on an InP substrate. The second is a 400 nm thick III-V heterostructure grown by MBE with a thin \textit{in-situ} deposited Al film. The last is a CPW on hte same wafer as that which is reported here, but with superconducting Al as the gate electrode as opposed to a combination of Cr and Au. The 400 buffer resonator was fabricated on the thin \textit{in-situ} aluminum on a similar layer structure as mentioned in this report, but with graded buffer layer steps of \SI{20}{\nano m} rather than \SI{50}{\nano m}, giving rise to a total thickness of approximately 400 nm. The Al growth conditions are nominally identical to that of the wafer presented previously. The InP wafer has 100 nm aluminum sputtered by DC magnetron sputtering after an Ar plasma cleaning in order to etch the native oxide. The design consists of a common feedline with hanger $\lambda/4$ CPW resonators with the same central conductor width and spacing to the ground plane as the device in the main text. The simulated $Q_\text{ext}$ for each CPW is 7830. The measured resonant frequency of the InP resonator is $f_r = \SI{7.717}{\giga Hz}$ and that of the 400 nm buffer resonator is $f_r = \SI{7.415}{\giga Hz}$.

\section{Appendix C: Measurement setup}

\begin{figure*}[htpb]
    \centering
    \includegraphics[width=0.8\textwidth]{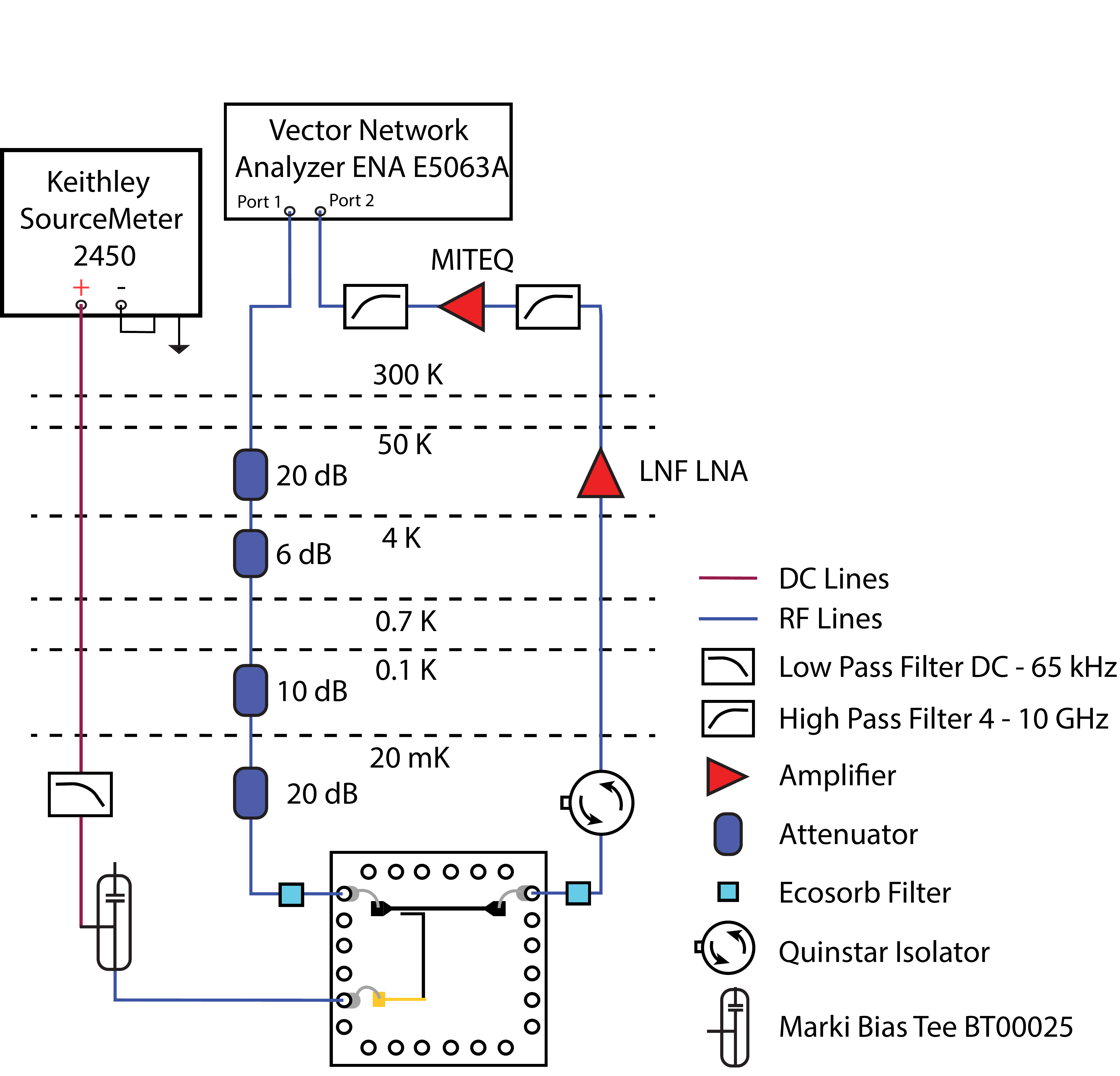}
    \caption{Measurement setup. A schematic of the cryogenic and room temperature measurement setup.}
    \label{fig:supp3}
\end{figure*}

A schematic of the measurement setup is shown in \cref{fig:supp3}. Measurements are conducted in an Oxford Triton dilution refrigerator. The sample is embedded in a QCage, a microwave sample holder manufactured by QDevil, and connected to the printed circuit board by aluminum wirebonds. Probe signals are sent from a vector network analyzer (Port 1) attenuated by -56 dBm with attenuation at each plate as noted. Attenuators are made by XMA. The signal then passes through a 1-18 GHz bandpass filter made by a copper box filled with cured Eccosorb castable epoxy resin. The signal is sent through the sample and returned through another Eccosorb filter, passed through an isolator with 20 dB isolation and 0.2 dB insertion loss, and then amplified with a low noise amplifier mounted to the 4K plate, as well as a room temperature amplifier (MITEQ) at room temperature. The gate electrode is connected to a voltage source and passed through a QFilter, a low pass filter manufactured by QDevil, mounted at the mixing chamber plate.

\section{Appendix D: Tunable resonator design}

\begin{figure}[ht]
    \centering
    \includegraphics[width=0.45\textwidth]{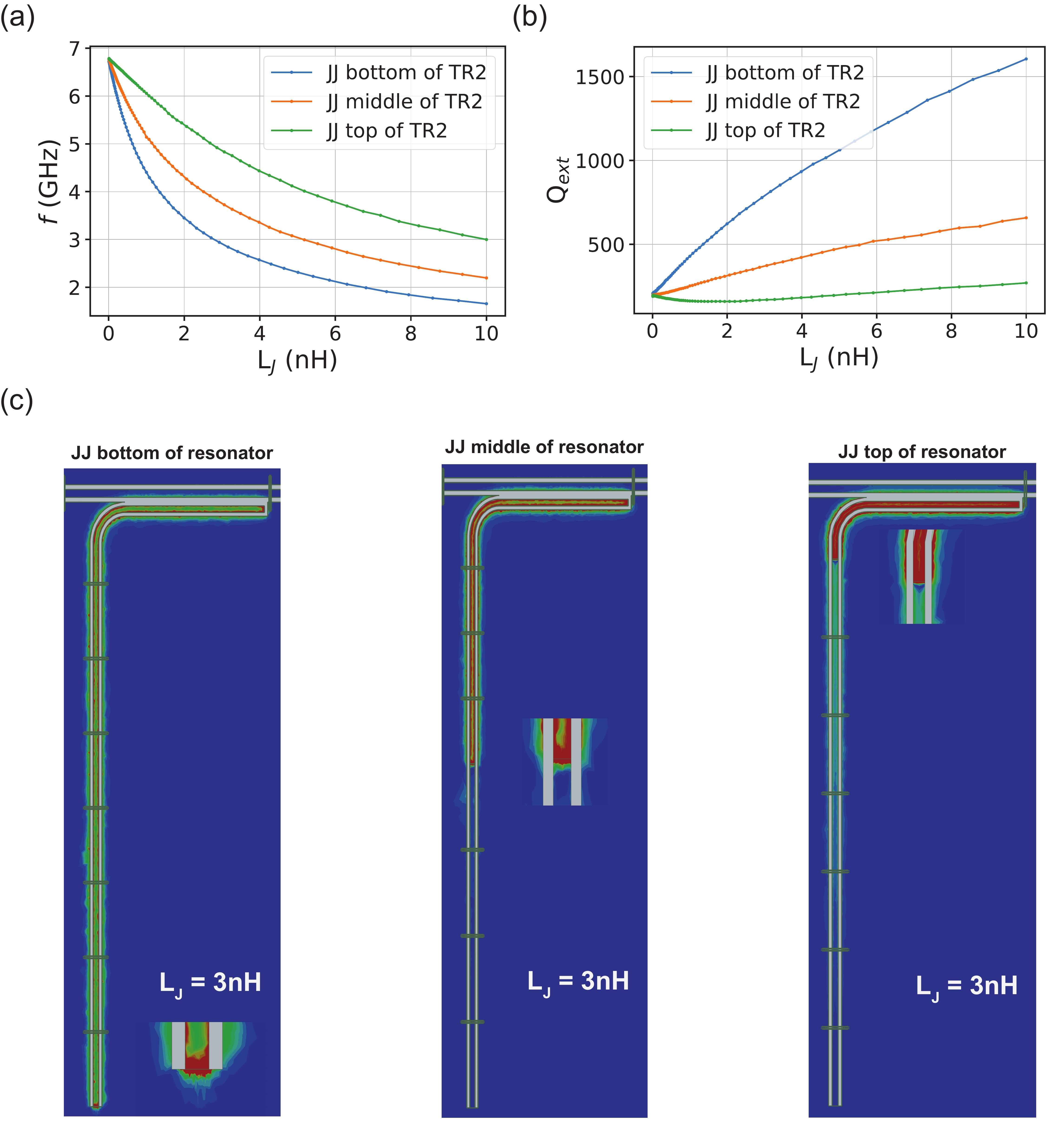}
    \caption{Effect of changing the position of the JJ in the tunable resonator. (a) Resonant frequency $f$ and (b) $Q_{ext}$ presented as a function of L$_{J}$ for three different cases: where the junction is at the top, middle and bottom of the resonator. (c) Finite-element calculations of the electric field distribution for each case with L$_{J} = $ \SI{3}{\nano H}. The inset shows a zoom in image of the junctions.}
    \label{fig:supp5}
\end{figure}

The details of the tunability the junction provides depends heavily on aspects of the design. The device presented in the manuscript has the JJ at the bottom of the resonator, directly shorting it to ground. Here, we also consider two other cases where the junction is in the middle and at the top of the resonator. As seen in \cref{fig:supp5}, having the junction at the bottom of the resonator provides slightly more tunability in $f$ while enhancing the value of $Q_\text{ext}$ compared to the two other cases. On the other hand, having the junction at the top (right before the bend) of the resonator provides slightly less tunability in $f$ while barely changing $Q_\text{ext}$. The range tunability of $Q_\text{ext}$ can be modified further by changing the initial value of $Q_\text{ext}$ and by switching the open and grounded ends of the CPW. 

\cref{fig:supp5}c presents the electric field profile of TR2 for the three cases with L$_{J} = $ \SI{3}{\nano H}. The distribution and concentration of the field can be seen to vary with the position of the junction. With the junction at the top of the resonator, the field can be seen to be restricted to the top part of the resonator while the rest of the resonator is isolated. This kind of isolation is ideal for superconducting quantum memory \cite{sardashti2020}.


\section{Appendix E: Critical power extraction}

\begin{table}
\centering
\begin{tabular}{|| c | c | c | c  || } 
 \hline
 $V_G$ (\SI{}{V})& $f_r$ (\SI{}{\giga Hz}) & $Q_L$ & $P_C$ (\SI{}{dBm})\\ [0.5ex] 
 \hline\hline
  5 & 6.114 & 435$\pm$5  & -64.3 \\ 
  0 & 6.114 & 473$\pm$2 & -64.5  \\ 
  -4 & 6.113 & 536$\pm$2  & -65.6 \\ 
  -8 & 6.100 & 497$\pm$1 & -64.7 \\
  \hline
\end{tabular}
\caption{Critical power extraction parameters: Summary of the parameters used to plot $f_B$ in terms of the reduced units $\Omega = 2Q_L(f_r-f_{B})/f_r$ and $P/P_C$. } 
\label{table:table}
\end{table}

In the bifurcation analysis, we find the critical power $P_{C}$ by analyzing the signal $S(f, P) = -|S_{21}(f, P)|$ which is a function of frequency $f$ and input power $P$. By discrete differentiation with respect to $f$ for fixed $P$, we get $\left( \partial S/ \partial f \right)_{P}$. The data is sampled with 1201 points in a span of \SI{100}{MHz}. The frequency at which the curve exhibits the highest derivative $\max(\partial S/ \partial f)_P$ is defined as $f_B$ and the value of the derivative at this point is $ \left(\partial S/ \partial f\right)_{f_B, P}$ which depends on the power $P$.

\begin{figure}
    \centering
    \includegraphics[width = .95\columnwidth]{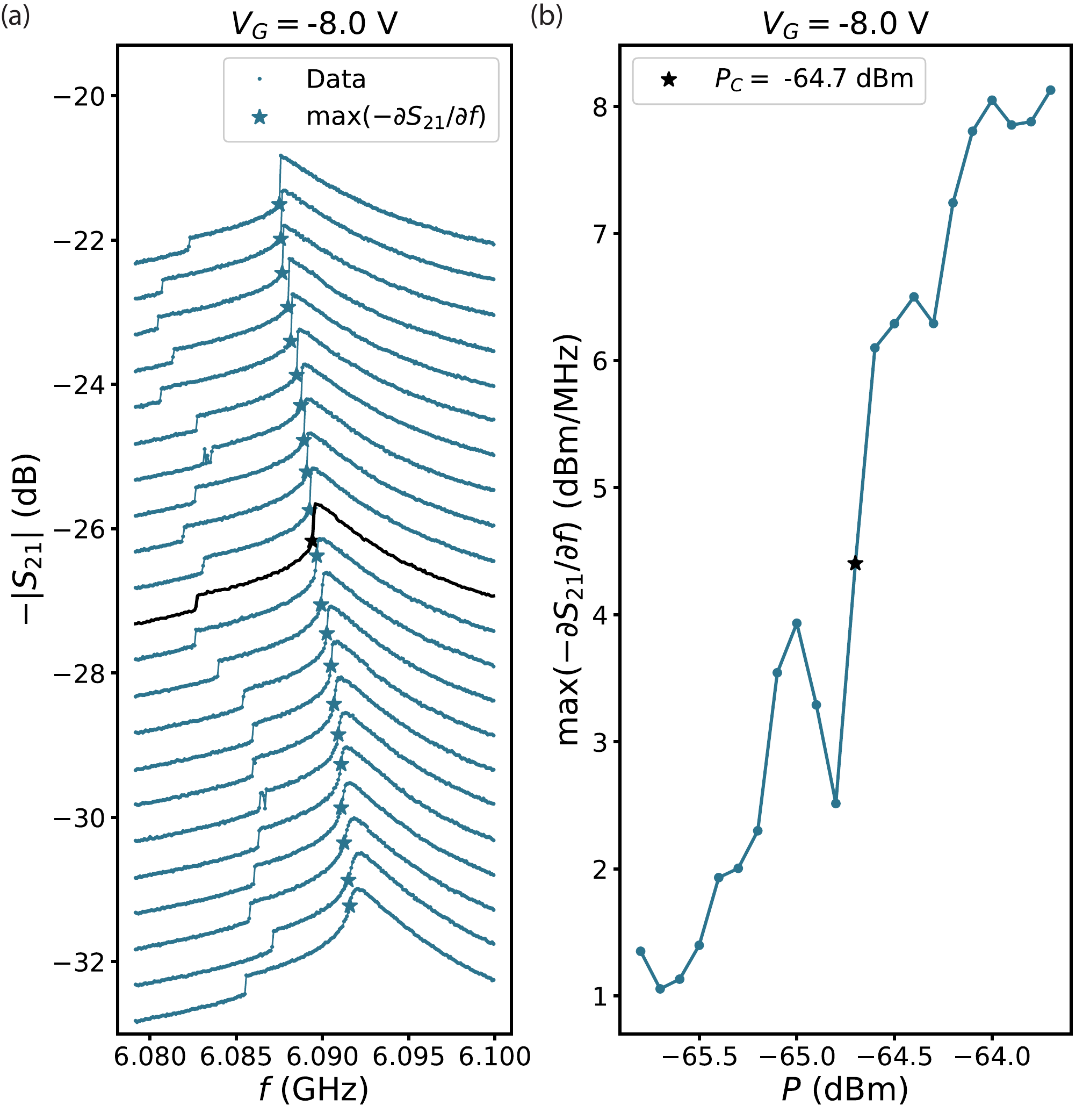}
    \caption{Extracting the critical power $P_C$. (a) Linecuts of $-S_{21}$ as a function of probe frequency $f$ plotted at different input powers $P$ with linecuts spaced for clarity. The top trace has $P=\SI{-56}{dBm}$ and the bottom trace has $P=\SI{-76}{dBm}$. (b) The derivatives of $\max(-\partial S_{21}/\partial f)$ evaluated as a function of $P$. The power at which this function is maximum is the critical power $P_C$, shown for this set of data as \SI{-64.7}{dBm}.}
    \label{fig:cp_extr}
\end{figure}

As $P$ increases, the critical point can be identified when $\partial S/\partial f$ first diverges. This divergence is due to the nature of the multi-valued solution of the Duffing equation which governs the systems behavior. When sweeping the frequency forward (from negative to positive), we find that the VNA probes only one solution branch at a time at powers greater than $P_C$. This gives rise to an apparent discontinuity in the data as seen in \cref{fig:cp_extr}a. We note that both solutions can be observed if one were to probe the system by sweeping frequency in the backward direction.

In order to identify the critical point, we look for the point at which the $(\partial S/\partial f)_{f_B, P}$ as a function of $P$ first . By taking a second derivative with respect to $P$, we identify at which power a sharp increase by finding the maximum, 

$$\max \left(\frac{ \partial }{ \partial P } \left[ \left(\frac{\partial S}{\partial f}\right)_{f_B, P} \right]\right).$$

The critical power $P_C$ is the power $P$ at which this occurs. Here, $\partial P= \SI{0.1}{dBm}$. An example of such extraction for $V_{G} = \SI{-8}{V}$ is presented in \cref{fig:cp_extr} and we summarize the various $P_C$ extracted for different gate voltages shown in \cref{table:table}. We also include the frequencies $f_r$ and loaded quality factor $Q_L$ used to rescale $f_B$ for \cref{fig:cp_extr}.

\section{Appendix F: Participation ratio calculations}

\begin{figure}
    \centering
    \includegraphics[width = .95\columnwidth]{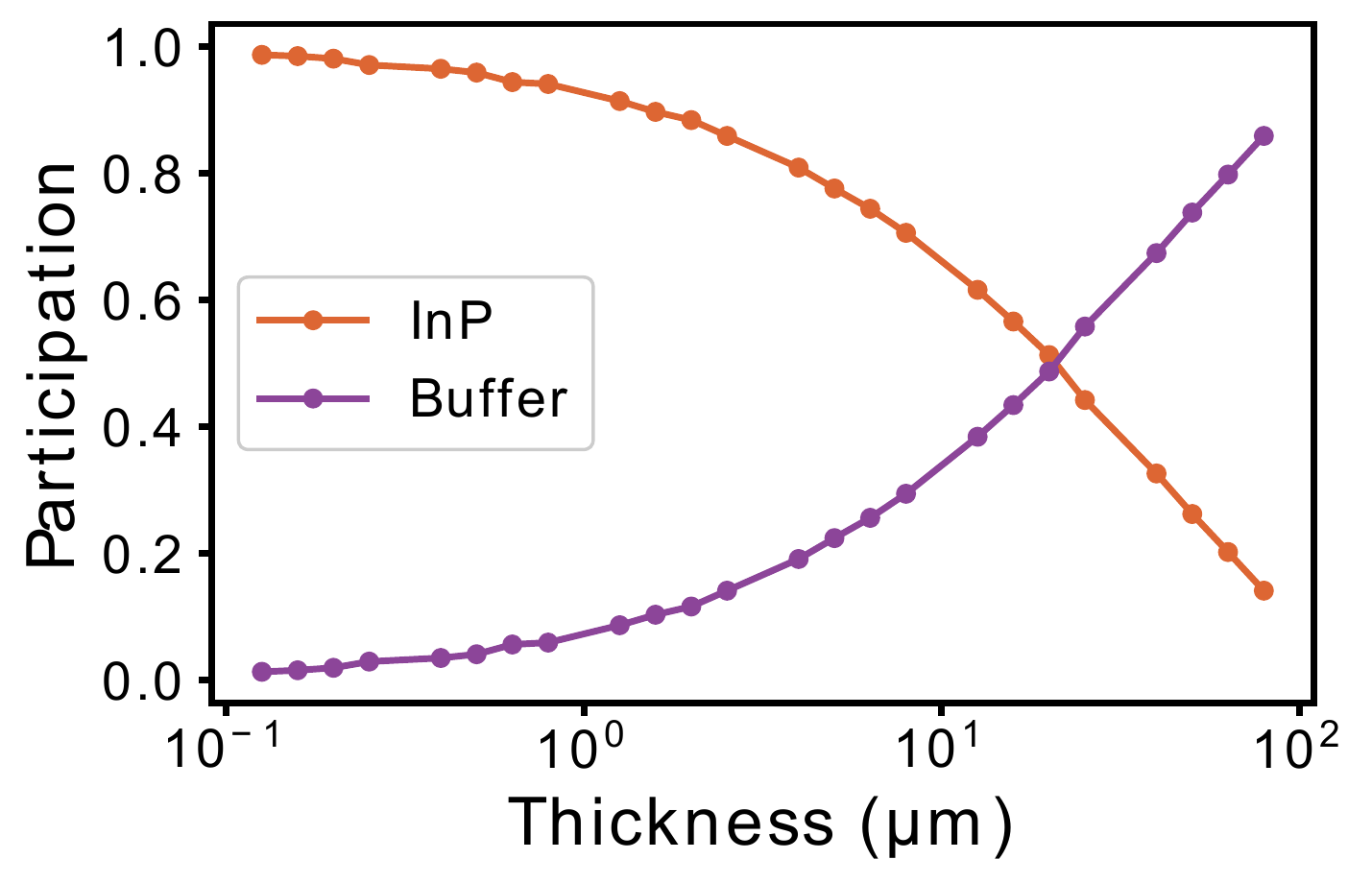}
    \caption{Participation ratio of buffer and InP as a function of buffer layer thickness.}
    \label{fig:parti}
\end{figure}

We further understand the buffer layer contributions to the loss by calculating the participation $p_i$ of the buffer layer and substrate as a function of buffer layer thickness using finite-element analysis (HFSS \cite{ansys}). The participation ratio is the ratio of the total electric field squared integrated over each volume $V_i$ 
$$p_i = \int_{V_i} \frac{|E^2|}{|E_\text{tot}^2|} dV.$$
Here $i$ only takes into account the \SI{500}{\micro m} InP substrate and the buffer layer. \cref{fig:parti} shows $p_i$ of the buffer layer and InP as a function of the thickness of the buffer layer. The participation ratio of the buffer layer is seen to increase with the thickness of the buffer laye which is expected.

\bibliography{References_Shabani_Growth}
\pagebreak

\end{document}